\documentstyle[aps,epsfig]{revtex}

\newcommand{\Va}{
$V_a$
}
\newcommand{\Vs}{
$V_s$
}

\begin{document}

\draft 

\title{Counting the Number of Bound States of Two-dimensional Screened Coulomb Potentials: A Semiclassical Approach}

\author{Christian Tanguy\footnote{phone: 33~1~42~31~78~81; fax: 33~1~42~53~49~30;\\ electronic address: christian.tanguy@rd.francetelecom.fr}}
\address{FRANCE TELECOM R\&D RTA/CDP, 196 avenue Henri Ravera, 92225 Bagneux Cedex, France
}
\date{\today}
\maketitle

\begin{abstract}
Portnoi and Galbraith recently proposed a beautiful and intriguing relationship defining the critical screening lengths associated with the apparition of new bound states for the two-dimensional statically screened Coulomb potential. Not only does semiclassical quantum theory show that this relationship is unfortunately not strictly exact, it has also proved helpful in the search for a potential which exactly verifies Portnoi and Galbraith's formula, namely $\displaystyle \frac{-e^2}{r \, (1+r/r_s)^2}$. The analytical eigenfunctions of the corresponding Schr\"{o}dinger equation at zero energy, when localized, lead to approximate upper bounds for critical screening lengths of the two-dimensional statically screened Coulomb potential.
\end{abstract}
\vskip0.3cm



\section{Introduction}

The two-dimensional statically screened Coulomb potential $V_s(r)$ plays a central role in the physics of semiconductor heterostructures near the band gap, especially with respect to their linear and nonlinear optical properties when the influence of excitons is properly taken into account\cite{Stern,Spector,Lee,Edelstein,HaugKoch,Xavier,Gubarev}. Although already considered three decades ago\cite{Stern}, it has nonetheless received a lot of recent attention and been analyzed using different approaches such as the WKB approximation\cite{Reyes}, perturbation theory\cite{Xavier}, variational calculation\cite{Spector,Lee,Edelstein} or numerical resolution based on the variable-phase method\cite{PG SSC,PG PRB58,PG PRB60}.

Using a variable-phase approach\cite{PG SSC} to count the bound states allowed by $V_s(r)$ --- only a finite number of them exist --- and making a connection with the Levinson's theorem in two dimensions, Portnoi and Galbraith recently proposed a beautiful and intriguing relationship between the critical screening lengths (scaled to the exciton Bohr radius) and the number of bound states for a given value of the angular momentum ($l \geq 0$):
\begin{equation}
(r_s)_{{}_{\scriptstyle c}} = \frac{(N+2\, l) \, (N+2\, l +1)}{2}  \hskip1cm (N \geq 0).
\label{relation PG}
\end{equation}
(the original notation used $\nu = N+1$, whereas we focus on the number $N$ of nodes of the wavefunctions). They proceeded to derive a formula giving the number of bound states as a function of $r_s$, which markedly differs from a WKB estimate\cite{Reyes} and from the Bargmann bound condition reformulated for the two-dimensional case\cite{PG PRB60}.

In this Communication, we show that semiclassical quantum theory simply demonstrates that Eq. (\ref{relation PG}) is not strictly exact for $l=0$ states, and that the previous estimate using the WKB approximation is incorrect. We then explain how this semiclassical approach has also proved useful in the search for a potential which obeys Eq.~(\ref{relation PG}), for which the eigenfunctions at zero energy are analytical. Finally, we show why, for the states defined by $N=0$ and $l \geq 2$, the true critical screening lengths of $V_s(r)$ are nearly given by Eq.~(\ref{relation PG}).

\section{Semiclassical quantum theory and the two-dimensional screened Coulomb potential}

The two-dimensional screened Coulomb potential $V_s$ has a very simple form when expressed in wavevector space\cite{HaugKoch,PG SSC}:
\begin{equation}
V_s(q) = - \frac{4 \, \pi}{q+q_s}, \hskip1cm \left( q_s \equiv \frac{1}{r_s} \right)
\label{Vs(q)}
\end{equation}
where we have adopted the same units as in Ref. \cite{PG SSC} (the length scale is the effective Bohr radius $a_0$ and the energy scale is the three-dimensional exciton energy $R$). In real space, it reads\cite{Stern,HaugKoch,PG SSC}
\begin{equation}
V_s(r) = - \frac{2}{r}  \Big\{ 1  - \frac{\pi}{2}  \, q_s \, r \, [{\mathbf H}_0(q_s \, r)-N_0(q_s \, r)]   \Big\}
\label{Vs(r)} \; ,
\end{equation}
where ${\mathbf H}_0$ and $N_0$ are the Struve and Neumann functions, respectively. The term between braces in Eq.~(\ref{Vs(r)}) goes down to 0 like $r_s^2/r^2$ as $r$ goes to infinity\cite{PG SSC}, which is a much reduced decrease with respect to the exponential decay occurring in three dimensions.

Let us now turn to the semiclassical quantum theory approximation for a potential $V(r)$. Using conformal mapping, Yi and collaborators have shown\cite{Yi} that the two-dimensional formulation of energy quantization may be written as
\begin{equation}
\int_{r_1}^{r_2} \sqrt{E - V_{\rm eff}(r)} \, dr = \left( N + \frac{1}{2} \right) \, \pi \; ,\hskip1cm(N \geq 0)
\label{semiclassique}
\end{equation}
where $V_{\rm eff}(r) = V(r) + l^2/r^2$, while $r_1$ and $r_2$ are the classical turning points. They have proved that Eq.~(\ref{semiclassique}) gives exact results in several cases, and excellent ones for the energy levels of impurity states in an arbitrary external magnetic field. Since we are looking for the critical values of $r_s$ at which new bound states appear {\em at zero energy}, we can set $E = 0$ in Eq.~(\ref{semiclassique}) so that, for the $l=0$ states of $V_s$,
\begin{equation}
\int_0^{+\infty} \sqrt{\frac{2}{r}  \Big\{ 1  - \frac{\pi}{2}  \, q_s \, r \, [{\mathbf H}_0(q_s \, r)-N_0(q_s \, r)] \Big\}} \, dr = \left( N + \frac{1}{2} \right) \, \pi \; .
\label{semiclassiqueStern}
\end{equation}
In the following we call $(\widetilde{r_s})_{{}_{\scriptstyle c}}$ the value obtained by using the semiclassical approximation in contrast to the exact one. Eq.~(\ref{semiclassiqueStern}) gives
\begin{equation}
(\widetilde{r_s})_{{}_{\scriptstyle c}} = \frac{1}{2} \, \left( \frac{\pi}{2 \, {\cal I}} \right)^2 \; \; \left( N + \frac{1}{2} \right)^2  
\label{semiclassiqueCT1}
\end{equation}
\begin{equation}
{\cal I} = \int_0^{+\infty} \sqrt{1  - \frac{\pi}{2}  u^2 \, [{\mathbf H}_0(u^2)-N_0(u^2)]} \, du .
\label{semiclassiqueCT2}
\end{equation}

As is usual with semiclassical expressions, we expect Eq.~(\ref{semiclassiqueCT1}) to be accurate when $N \gg 1$. If Eq.~(\ref{relation PG}) were true, we would find $2 \, {\cal I} = \pi$, so as to get the correct leading dependence of $(r_s)_{{}_{\scriptstyle c}}$ in $N^2$. ${\cal I}$ has been numerically evaluated by first calculating the integral between 0 and 10, and then bracketing the integral from 10 to $+\infty$ by using the asymptotic expansion of ${\mathbf H}_0-N_0$. The same procedure, repeated for an intermediate bound of 20, gave the same result, namely
\begin{equation}
2 \, {\cal I} \approx 3.14057414.
\label{valeur numérique de I}
\end{equation}
It definitely differs from $\pi$, even though the relative error is only about $3.24 \times 10^{-4}$. Consequently, for $l=0$ states and large $N$'s, $(r_s)_{{}_{\scriptstyle c}}$ should be different from Eq.~(\ref{relation PG}).

Let us now discuss previously published results in the light of Eqs.~(\ref{semiclassiqueCT1}) and (\ref{valeur numérique de I}). Portnoi and Galbraith \cite{PG PRB58,PG PRB60} compared Eq.~(\ref{relation PG}) with the WKB prediction of Reyes and del Castillo-Mussot for $l=0$ states\cite{Reyes}, and observed a disagreement by a factor 2.5. By contrast, Eq.~(\ref{valeur numérique de I}) satisfactorily removes the discrepancy. Why then do our semiclassical approximation and the WKB result differ by such a large amount ? The explanation probably comes from the tricky use of WKB wavefunctions and asymptotic expansions in Eqs.~(7)-(12) of Ref.~\cite{Reyes}. In particular, their Eq.~(12) contains $\int_0^1 w^{-7/6} \, \sqrt{1-w} \, dw$, which diverges. Admittedly, this integral can be {\em formally} written as $B(3/2,-1/6)$, whose true value is -6.72 instead of the given -2.81. This oversight seems to have been left uncorrected until the final result in their Eq.~(18), casting therefore strong doubt about its validity.

At this point, we have merely shown that Portnoi and Galbraith's formula cannot be exact for $l=0$ states, since it does not verify the large $N$ limit. Strictly speaking, we cannot use this argument to reject their hypothesis that $(r_s)_{{}_{\scriptstyle c}} = $1, 3, 6, 10, ... for $l=0$ and $N$ = 1, 2, 3, 4 ... or more generally, for low-energy states. However, because of the near-coincidence of $2 \, {\cal I}$ with $\pi$, one may have second thoughts about the accuracy of their $(r_s)_{{}_{\scriptstyle c}}$: the statically screened Coulomb potential is very long range because of its $1/r^3$ decrease at infinity\cite{condition initiale}. Besides, a question remains unanswered: is it possible to understand why the critical screening lengths are so close to the integers of Eq.~(\ref{relation PG})? For this reason, we investigated what happens if $V_s$ is replaced by $\displaystyle - \frac{2}{r} \, f(\frac{r}{r_s})$, when $f \rightarrow 1$ for $r \rightarrow 0$ and $f(x)$ monotonically decreases as $1/x^2$ for $x \rightarrow \infty$.

\section{Semiclassical approximation for a family of screened Coulomb potentials}

For an attractive potential $\displaystyle - \frac{2}{r} \, f(\frac{r}{r_s})$, Eq.~(\ref{semiclassique}) written at zero energy gives
\begin{equation}
(\widetilde{r_s})_{{}_{\scriptstyle c}} = \frac{\pi^2}{\displaystyle 8 \left[ \int_0^{+\infty} \sqrt{f(x^2)} \; dx \right]^2} \; \left( N + \frac{1}{2} \right)^2 . 
\label{semiclassiquepot1}
\end{equation}
Let us call ${\cal A}$ the prefactor of $(N+1/2)^2$. We see on Table~\ref{table des constantes pour les potentiels écrantés} that even though the asymptotic behaviors are the same for all the potentials ($-2/r$ and $-2 \, r_s^2/r^3$ for $r$ going to 0 and $\infty$, respectively), ${\cal A}$ may vary between at least 1/3 and 2/3. The criterion of apparition of bound states is thus extremely sensitive to the exact shape of the potential over its whole range of variation. Obviously, for
\begin{equation}
V_a(r) = - \frac{2}{r}  \, \frac{1}{\displaystyle \left( 1+ \frac{r}{r_s} \right)^2} \; ,
\label{Va(r)}
\end{equation}
we have ${\cal A} = 1/2$. Because of its simple form, one can further investigate this potential for $l \neq 0$. The calculation is performed without difficulty and yields
\begin{equation}
(\widetilde{r_s})_{{}_{\scriptstyle c}} = \frac{(N+2\, l+\frac{1}{2})^2}{2}  \hskip1cm (N,l \geq 0).
\label{semiclassique pour Va}
\end{equation}
Equation~(\ref{semiclassique pour Va}) strikingly resembles Eq.~(\ref{relation PG}), apart from a difference of 1/8. Since simple results derived from semiclassical quantum theory are often closely related to solvable problems, we have searched for the exact analytical expression of the critical screening lengths for $V_a$ given in Eq.~(\ref{Va(r)}).

\section{Exact critical screening lengths for \Va}

For the potential $V_a$, the Schr\"{o}dinger equation verified by the radial wavefunctions $\psi_0(r)$ of zero energy reads
\begin{equation}
\psi_0''(r)+ \frac{1}{r} \, \psi_0'(r) + \Bigg( \frac{2}{r}  \, \frac{r_s^2}{(r_s+ r)^2} -\frac{l^2}{r^2} \Bigg) \, \psi_0(r)= 0.
\label{ODE pour Va}
\end{equation}
Because of the particular form of $V_a$, one may first consider rational fractions of $r$ for $\psi_0$. Starting with $N=0$ (for wavefunctions having no node), one can check that $r^l/(r+r_s)^{2 \, l}$ obeys Eq.~(\ref{ODE pour Va}) provided that $r_s = l \, (2 \, l+1)$. Proceeding further with $N=1,2, ...$, it is not difficult to see that the structure of the solutions is
\begin{equation}
\psi_0(r) \; \propto \; \frac{r^l}{\Big(r+\frac{(N+2\, l) \, (N+2\, l +1)}{2}\Big)^{N+2 \, l}} \; \; {}_2F_1\Bigg(-N,-N-2\, l;2 \, l+1;- \frac{2 \, r}{(N+2\, l) \, (N+2\, l +1)}\Bigg) \; ,
\label{Psi critique 1}
\end{equation}
where ${}_2F_1$ is the hypergeometric function, which here reduces to a polynomial of degree $N$. Quite remarkably, the critical screening lengths are exactly given by Portnoi and Galbraith's formula, Eq.~(\ref{relation PG}). This also shows a posteriori that semiclassical quantum theory can be trusted to provide the correct result when $N$ is large.

A more direct way to solve Eq.~(\ref{ODE pour Va}) is to set $\displaystyle \psi_0(r) = r^l \, \chi(\frac{r}{r+r_s})$; $\chi(u)$ then satisfies the hypergeometric equation. Eq.~(\ref{relation PG}) is obtained through the condition that the radial wavefunction, which is finite at the origin, does not diverge at infinity: one of the first two arguments of ${}_2F_1$ must then be a nonpositive integer. This leads to
\begin{equation}
\psi_0(r) \; \propto \; r^l \; \; {}_2F_1\Bigg(-N-2\, l,N+2 \, l+1;2 \, l+1;\frac{r}{r+\frac{(N+2\, l) \, (N+2\, l +1)}{2}}\Bigg) \; ,
\label{Psi critique 2}
\end{equation}
which is equivalent to Eq.~(\ref{Psi critique 1})\cite{GR}. These wavefunctions are currently considered for the determination of upper bounds for the critical screening lengths of $V_s$, when $l \geq 1$\cite{bibi}.

\section{Upper bounds for critical screening lengths of \Vs}

The above wavefunctions have an interesting property for $l \geq 2$, which is obvious from Eq.~(\ref{Psi critique 1}): they are localized (square integrable). Since $V_a$ and $V_s$ have the same asymptotic behaviors and seem to have nearly identical $(r_s)_{{}_{\scriptstyle c}}$, we may be tempted to use perturbation theory to get an {\em estimate} of the critical screening lengths of $V_s$. In the following, we will restrict the discussion to the ($N=0$, $l \geq 2$) states.

Starting from the exact solution $\psi_0$ of zero energy given in Eq.~(\ref{Psi critique 1}), the first-order correction to the energy is obtained by calculating the expectation value of $\Delta V = V_s-V_a$ on this state, namely
\begin{equation}
\Delta E^{(1)}_{N=0,l \geq 2} =  \langle \psi_0 | \; \; V_s(r) +\frac{2}{r} \, \frac{l^2 \, (2 \, l+1)^2}{\big( r+l \, (2 \, l+1) \big)^2} \; \; | \psi_0 \rangle \; .
\label{DeltaE 1 de base}
\end{equation}
Assuming that $(r_s)_{{}_{\scriptstyle c}} = l \, (2 \, l+1)$ for $V_s$ leads to (after normalization of $\psi_0$)
\begin{eqnarray}
\Delta E_{N=0,l \geq 2}^{(1)} & = & - \frac{2 \, (4 \, l-1)!}{l\, (2 \, l+1) \, (2 \, l+1)! \, (2 \, l-3)!} \nonumber \\
& & \times \int_0^{+\infty} \, du \, \frac{u^{2 \, l}}{(1+u)^{4 \, l}} \, \bigg( 1 - \frac{\pi}{2} \, u \, [{\mathbf H}_0(u)-N_0(u)] -\frac{1}{(1+u)^2} \bigg) \; .
\label{Delta E ordre 1}
\end{eqnarray}
The above integral is much easier to evaluate numerically than ${\cal I}$, because $m! \, (1+u)^{-m-1}$ is the Laplace transform of $\displaystyle t^{m} \, e^{-t}$ and because\cite{GRbis}
\begin{equation}
\int_0^{+\infty} du \, \frac{\pi}{2} \, [{\mathbf H}_0(u)-N_0(u)] \, e^{- u \, t} = \frac{1}{\sqrt{1+t^2}} \, \ln \Big( \frac{(1+\sqrt{1+t^2}) \, (t+\sqrt{1+t^2})}{t} \Big) \; .
\label{GRCT}
\end{equation}
One finds that $\Delta E^{(1)}_{N=0,l \geq 2}$ is equal to $1.29 \times 10^{-4}$, $1.22 \times 10^{-4}$, $9.87 \times 10^{-5}$, $7.83 \times 10^{-5}$, $6.27 \times 10^{-5}$, $5.10 \times 10^{-5}$, $4.22 \times 10^{-5}$ for $l =2, ..., 8$, respectively. These small positive values seem to indicate that for the lowest-lying states with $l \geq 2$, the critical screening lengths of the two-dimensional statically screened Coulomb potential are probably larger than $l \, (2 \, l+1)$, but only slightly so.

A visual way to reach this conclusion is to consider the different functions appearing in the integral of Eq.~(\ref{Delta E ordre 1}). We see in Fig.~\ref{Fig1} that $r \, \Delta V(r)$ goes to zero at the origin and $r \rightarrow \infty$, as expected; however, its {\em sign} is not uniform. In order to get $\Delta E^{(1)}_{N=0,l \geq 2}$, one must also consider the squared wavefunctions, $|\psi_0|^2$, the amplitude of which rapidly decreases with $l$. Furthermore, the maximum of $|\psi_0|^2$ is always located at $u=1$, i.e., $r = l \, (2 \, l+1)$; obviously, positive and negative contributions to the integral nearly compensate each other. Since the first-order energy corrections are very small already, we can understand why the critical screening lengths for $V_s$ must be very close to $l \, (2 \, l+1)$ for the ($N=0$, $l \geq 2$) states. 

For these states, we can actually use second-order perturbation theory to get an admittedly rough upper limit to the exact value of $(r_s)_{{}_{\scriptstyle c}}$ for $V_s$. The argument is the following: it is well known that second-order energy corrections to the state of lowest energy are {\em always} negative. This result can be extended here to the $N=0$ states, since subspaces corresponding to different $l$'s remain uncoupled by the isotropic $\Delta V$. If, instead of taking $r_s = l \, (2 \, l+1)$ as in Eq.~(\ref{Delta E ordre 1}), one considers $r_s =  l \, (2 \, l+1)/\eta_l$ with $\eta_l$ {\em such that} $\Delta E^{(1)}_{N=0,l \geq 2} = 0$, then $\Delta E^{(2)}_{N=0,l \geq 2} < 0$, which suggests the possible existence of another bound state and consequently $(r_s)_{{}_{\scriptstyle c}} < l \, (2 \, l+1)/\eta_l$. As regards calculations, one merely has to replace all the $u$'s in the $V_s$ part of Eq.~(\ref{Delta E ordre 1}) by $\eta_l \, u$. Finally, one gets $(r_s)_{{}_{\scriptstyle c}}^{[l=2]} < 10.106$, $(r_s)_{{}_{\scriptstyle c}}^{[l=3]} < 21.273$, $(r_s)_{{}_{\scriptstyle c}}^{[l=4]} < 36.524$. Although rather crude, these upper bounds have been verified (and improved) by a more accurate variational approach, which also showed that $(r_s)_{{}_{\scriptstyle c}}^{[N=1,l=1]}$ is strictly less than 6\cite{bibi}.

\section{Conclusion}

In conclusion, we have shown that Eq.~(\ref{relation PG}), proposed by Portnoi and Galbraith to describe the critical screening lengths for the two-dimensional statically screened Coulomb potential $V_s$, is unfortunately not strictly exact. This does not invalidate their variable-phase method approach, which is very useful: it is quite likely that the discrepancy between the exact values and the integers they found must be very small. Using analytical eigenfunctions at zero energy for a potential very similar to $V_s$, we have been able to explain why the critical screening lengths corresponding to the fundamental states for $l \geq 2$ are indeed nearly given by Eq.~(\ref{relation PG}).

\section*{Acknowledgements}

It is a pleasure to thank M. Combescot, F. Glas, L. Largeau, G. Patriarche, and J. Vigu\'{e} for very useful discussions and suggestions.


\begin{table}

\vskip1.0cm

\caption{Values of ${\cal A}$ such that $\displaystyle (\widetilde{r_s})_{{}_{\scriptstyle c}} = {\cal A} \, (N+\frac{1}{2})^2$ in the semiclassical approximation,  for different potentials $\displaystyle \frac{- e^2}{r} \, f(\frac{r}{r_s})$.}
\label{table des constantes pour les potentiels écrantés}

\vskip1.0cm

\begin{tabular}{|cl|} 
$f(x)$ & ${\cal A}$ \\[0.1cm] \hline 
$1  - \frac{\pi}{2} \, x \, [{\mathbf H}_0(x)-N_0(x)]$ \hskip1cm & $\approx 0.500324$ \\[3mm] 
$\displaystyle \frac{1}{1+x^2}$ & $\displaystyle \frac{2 \, \pi^3}{\Gamma(\frac{1}{4})^4} \approx 0.358885$\\[3mm]
$\displaystyle \frac{(1-\exp(-\sqrt{x}))^4}{x^2}$ & $ \displaystyle \frac{\pi^2}{32 \, (\ln 2)^2} \approx 0.641947$\\[3mm]
$\displaystyle \frac{(1-\exp(-x))^2}{x^2}$ & $\displaystyle \frac{\pi}{8} \approx 0.392699$\\[3mm]
$\displaystyle \frac{{\rm tanh}^2(x)}{x^2}$ & $\displaystyle \frac{(9+4 \, \sqrt{2}) \, \pi^3}{196 \, \zeta(\frac{3}{2})^2} \approx 0.339753$\\[3mm]
$\displaystyle \frac{{\rm tanh}^4(\sqrt{x})}{x^2}$ & $\displaystyle \frac{\pi^6}{1568 \, \zeta(3)^2} \approx 0.424329$\\[3mm]
$\displaystyle \frac{1}{(1+x)^2}$ & $\displaystyle \frac{1}{2}$
\end{tabular}
\end{table}


\begin{figure}
\caption{
Behavior of $\displaystyle 1 - \frac{\pi}{2} \, u \, [{\mathbf H}_0(u)-N_0(u)] -\frac{1}{(1+u)^2}$ (full line) and of $\displaystyle \frac{u^{2 \, l}}{(1+u)^{4 \, l}}$, which are the squares of the wavefunctions used in Eq.~(\protect\ref{Delta E ordre 1}). For clarity, the curves for $l=2, 3, 4$ have been multiplied by 10, 100, and 1000, respectively.
}
\label{Fig1}
\end{figure}

\vspace*{-6.5cm}
\hskip-2cm
\includegraphics{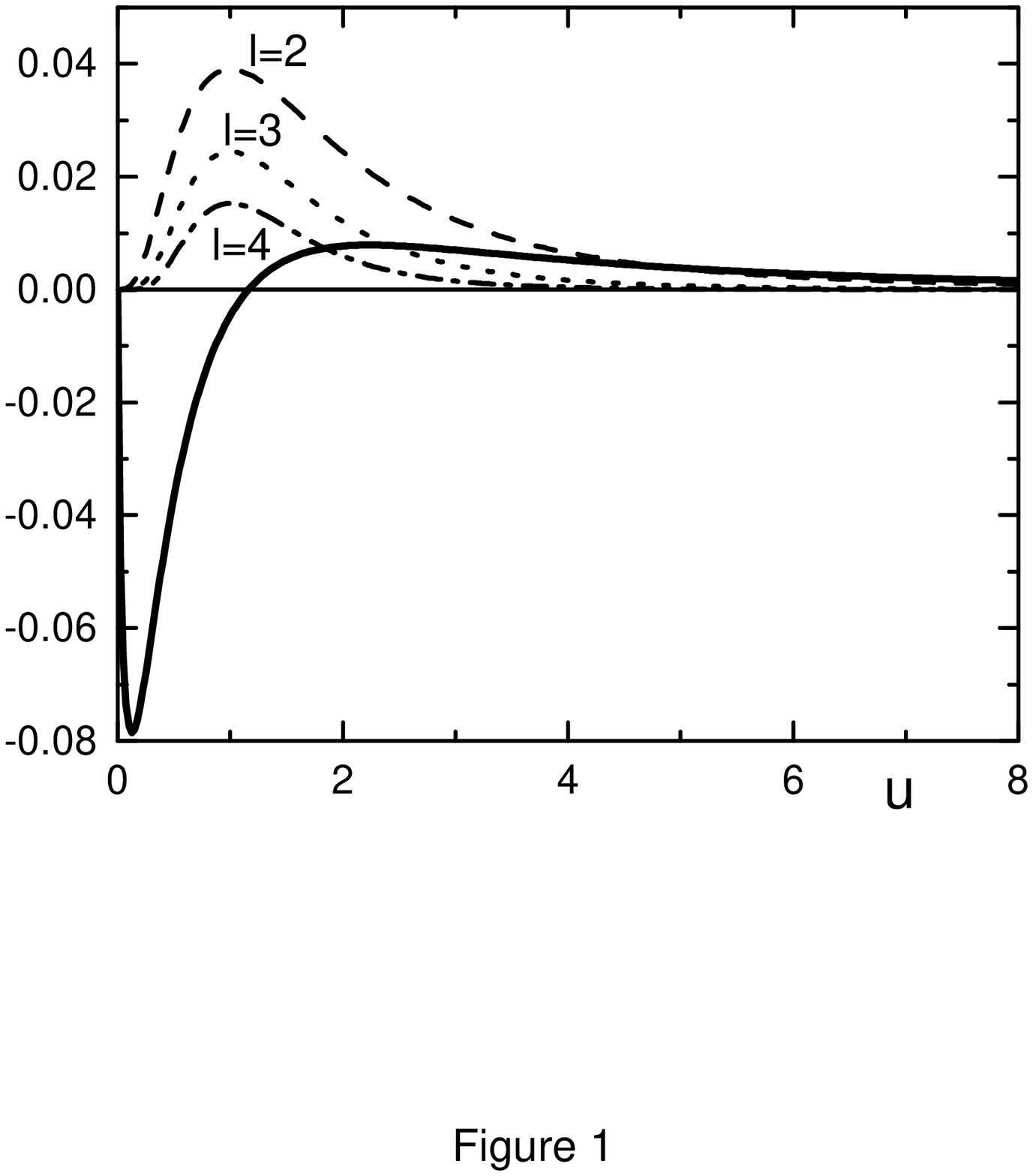}


\begin{references}
\bibitem {condition initiale} When solving the differential equations for $\delta_0(\rho)$ and $\mu_0(\rho,\kappa)$, Portnoi and Galbraith adopt boundary conditions for $\rho_0 > 0$, which do not exactly agree with those obtained by expanding the functions as double series in powers of $\rho$ and $\ln(\rho \, \exp({\mathbf C})/2)$.
\end{references}
\end{document}